\documentclass{article}

\usepackage{arxiv}

\usepackage[utf8]{inputenc} 
\usepackage[T1]{fontenc}    
\usepackage{hyperref}       
\usepackage{url}            
\usepackage{booktabs}       
\usepackage{amsfonts}       
\usepackage{nicefrac}       
\usepackage{microtype}      
\usepackage{lipsum}		
\usepackage{graphicx}
\usepackage{doi}
\usepackage{amsmath}

\title{Investigating thermal transport in knotted graphene nanoribbons using non-equilibrium molecular dynamics}

\author{ 
    \hspace{1mm}Levi C.~Felix, Alexandre F. Fonseca and Douglas S.~Galvao\thanks{Corresponding author: galvao@ifi.unicamp.br} \\
	Department of Applied Physics\\
	State University of Campinas\\
	Campinas, SP 13083-970, Brazil \\
}

\renewcommand{\shorttitle}{Thermal Transport in Knotted GNRs}

\begin{document}
\maketitle

\begin{abstract}
    In this work, we investigated the effect of knots in the thermal transport of graphene nanoribbons through non-equilibrium molecular dynamics simulations. We considered the cases of one, two, and three knots are present. Temperature jumps appear in the temperature profile where the knots are located, which indicates that they introduce thermal resistances in the system, similar to interfacial Kapitza resistance present between two different materials and/or single materials with defects and/or lattice distortions. We found that the thermal resistance introduced by each individual knot is essentially the same as the overall resistance increase linearly with the number of knots, as they behave as thermal resistances associated in series. Also, the relative position between each knot in the arrangement does not strongly affect the thermal current produced by the temperature gradient, showing a weak thermal rectification effect.
\end{abstract}

\keywords{Graphene nanoribbons \and Thermal transport \and Non-equilibrium molecular dynamics}

\section{Introduction}

Among all graphene-derived nanostructures, graphene nanoribbons (GNRs) have received much attention from the materials physics community due to their quasi-one-dimensional character and topological simplicity, as they are only narrow and straight-edged stripes of graphene. Particularly, the electronic properties have been found to strongly depend on the edge type (armchair or zigzag)~\cite{wakabayashi_2010,son_2006, nakada_1996,jiang_2008} where localized states appear at the edges of zigzag GNRs but not on zigzag ones. More recently, even some topological phases under external electric fields and boron/nitrogen doping have been reported~\cite{zhao_2021}. This creates perspectives for GNRs to be a competitive candidate material in quantum electronics applications~\cite{wang_2021}. For such applications, it is also important to study their thermal transport since there is always a heating-up/cooling-down cycle in electronic devices. For most low-dimensional materials, thermal transport is expected to be influenced by surface roughness where the majority of phonon scattering processes take place~\cite{aksamija_2012,fugallo_2018}. For GNRs, however, the mean free path of the dominant phonon mode ($\sim 1$~nm) is much larger than the mean height variation at the surface ($0.2-0.6$~\AA)~\cite{wang_2012}. The successful experimental synthesis of GNRs~\cite{cai_2010,li_2008,jiao_2009} has made possible the investigation of their electronic~\cite{wang_2008} and thermal~\cite{chen_2022} transport properties, which opens many possibilities for industrial large-scale applications. 

Defect engineering of GNRs is a common strategy to improve and tune some properties that otherwise are not observed in non-defective nanoribbons. Examples include spin filtering by combining vacancies and doping with boron and nitrogen atoms~\cite{mandal_2014}, and bandgap tuning through edge patterning by interacting with vacancy defects~\cite{du_2017}. Geometrical distortions, such as twists and knots, have also been considered to obtain new behavior in addition to point and edge defects. The possible formation of twisted GNRs was already discussed~\cite{nikiforov_2014,fonseca_2021}, where a helix configuration was found to be more favorable than a simple torsion. Also, the charge carrier type and mobility could also be tuned by the degree of twisting and turning on GNRs helices~\cite{thakur_2020}. Other interesting findings on twisted GNRs include unique properties for different conformations of the nanoribbons~\cite{shahabi_2014}, presence of non-linear current-voltage characteristics with an applied transverse electric field~\cite{saizbretn_2019}, structural strengthening effect when they are inside carbon nanotubes~\cite{fang_2016}, and even chirality-controlled carbon nanotube synthesis~\cite{kit_2012}. Knots are also another sort of material distortion that have been investigated in carbon materials like carbon nanotubes~\cite{rego_2019}, microfibers~\cite{bosia_2016,pugno_2014}, graphene oxides~\cite{xiang_2013}, silk fibers~\cite{pantano_2016}, and GNRs~\cite{kagimura_2012}. Particularly, knotted GNRs were found to possess two metastable configurations at two distinct stretching levels and a spin-polarized electronic ground state in the tight-knot case.

Extensive mechanical properties studies in knotted nanostructures have already been reported in the literature~\cite{rego_2019,bosia_2016,pugno_2014,xiang_2013,pantano_2016,kagimura_2012}, but very few have been focused on investigating heat transport. In this way, in the present work, we investigated the influence of knots GNRs on the heat flux generated by the presence of two thermal contacts at distinct temperatures that creates a temperature gradient. For this purpose, we considered the presence of one, two, and three knots. For direct comparison, we have also considered the cases of the structures without defects nor twists, to better evaluate the role played by such distortions on the thermal transport of knotted GNRs.

\section{Methods}

\begin{figure}[!htb]
\centering
\includegraphics[width=\linewidth]{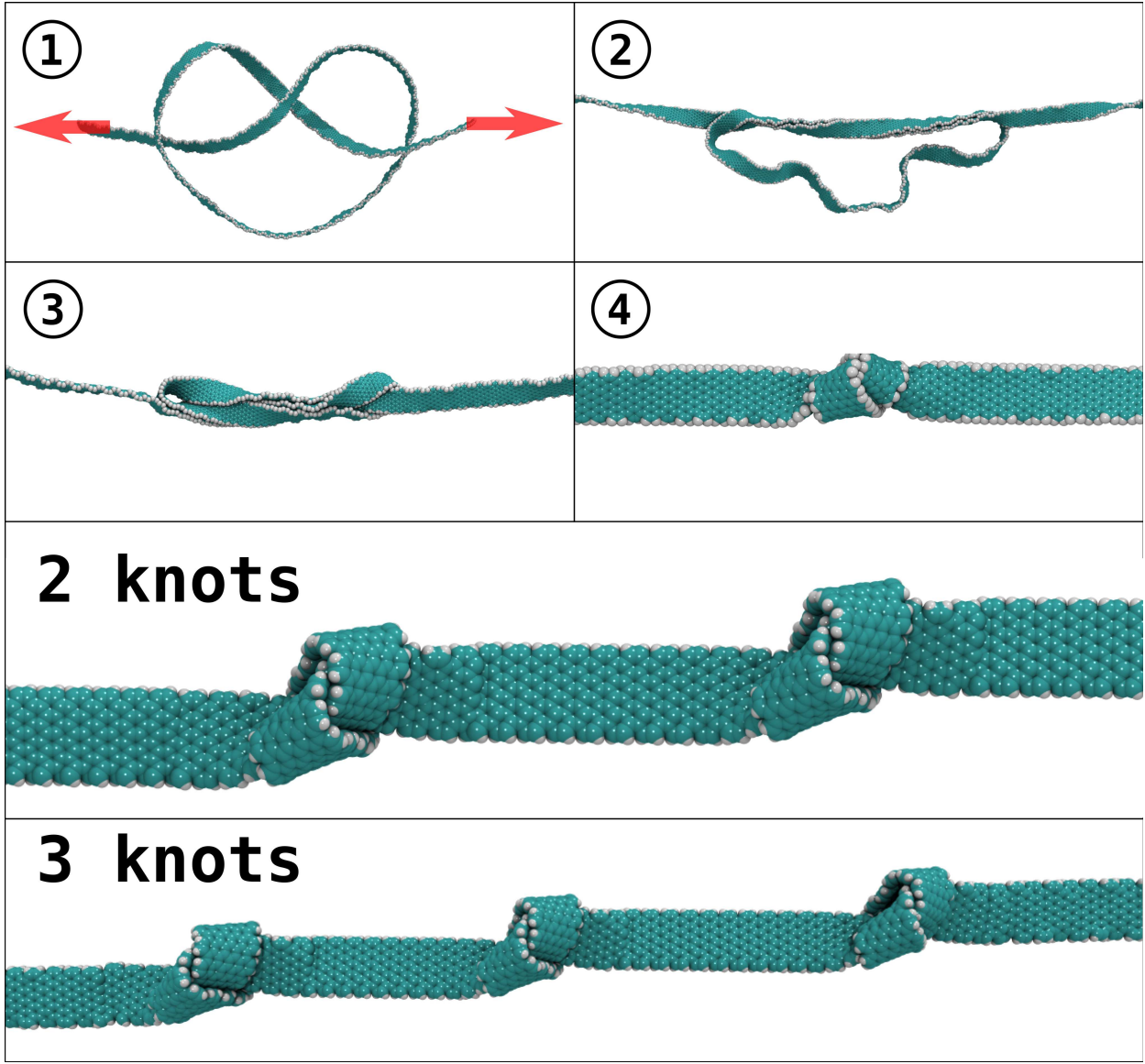}
\caption{Sequential snapshots showing the process of tying the knot in hydrogen passivated GNR (1-4). Two and three-knot structures are shown at the bottom, where each knot is 60~\AA~apart. Green and white colors indicate carbon and hydrogen atoms, respectively.}
\label{fig:scheme}
\end{figure}

To build a knotted hydrogen passivated GNR, we created a straight structure with width of $W=10$~\AA~ using the Visual Molecular Dynamics (VMD) software. Then, with the help of the sculptor tool, we considered a set of four points where the nanoribbon is mapped to obtain a pre-knotted structure that is shown in the snapshot 1 of the Figure \ref{fig:scheme}. Then, using fully atomistic reactive molecular dynamics simulations, we stretched the whole structure until the knot is closed, as shown in the snapshots 2-4 of Figure \ref{fig:scheme}. We repeated the same process to create the two and three-knot GNRs shown in Figure \ref{fig:scheme} at the bottom. The distance between each knot was chosen to be $d=60$~\AA~ and the total length $L=3$~nm. As mentioned above, for comparison we also considered a GNR without knots with the same length.

\begin{figure}[!htb]
\centering
\includegraphics[width=\linewidth]{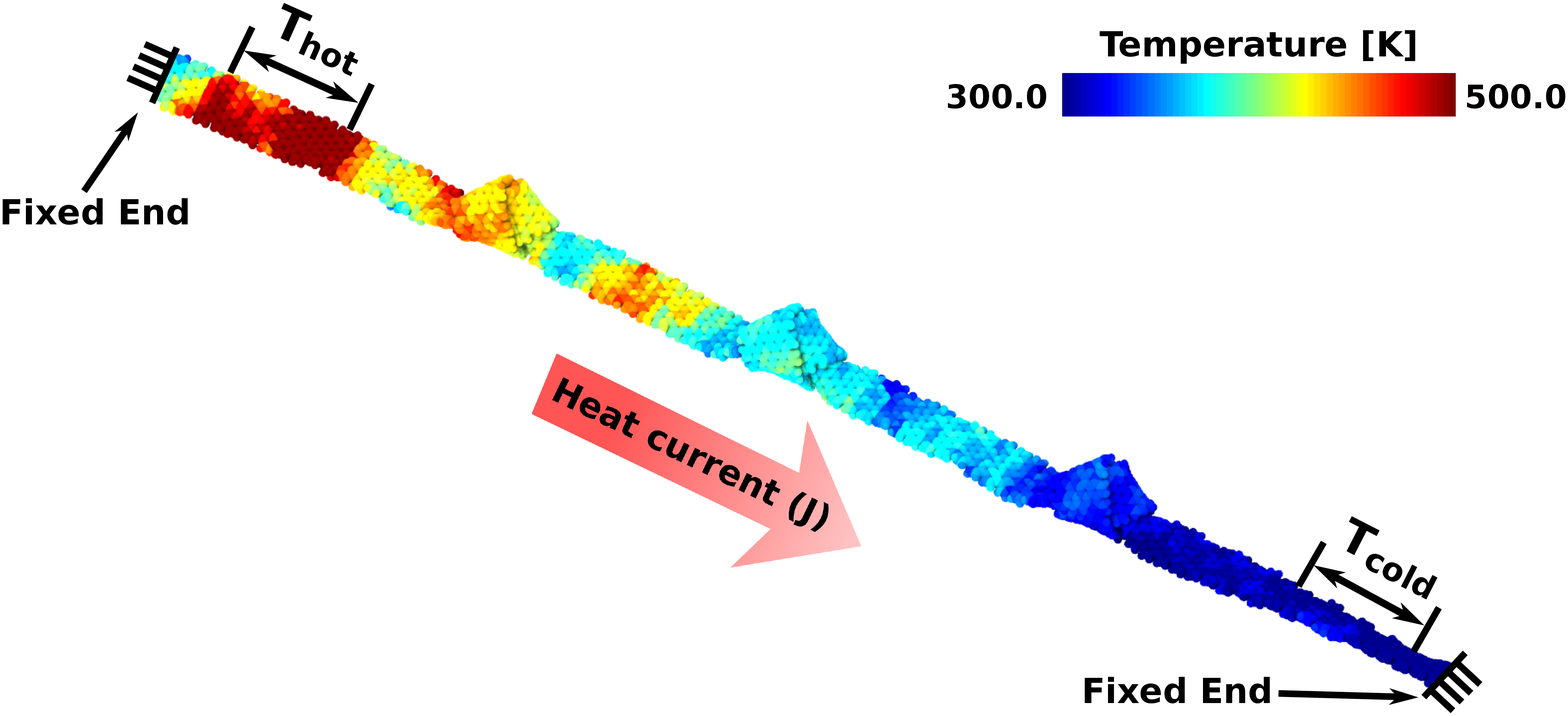}
\caption{Scheme of the simulation setup of generating a temperature gradient in a knotted GNR. Both GNR ends are fixed and the heat baths are indicated for the hot and cold reservoir. Atoms are colored accordingly to their spatially-averaged atomic temperature $t$, where the average considers for all atoms within a cutoff distance of 10~\AA to the central atom. The temperature color bar scale is shown at the upper right.}
\label{fig:temperature}
\end{figure}

We considered all nanoribbons in this work to be suspended by fixing both of their ends, as illustrated by Figure \ref{fig:temperature}. We investigated the thermal transport of knotted GNRs through molecular dynamics (MD) simulations using the LAMMPS code~\cite{plimpton_1995}. We first equilibrated all GNRs using an NVT ensemble during 100~ps to determine their structural and thermal stability. The interaction among all carbon atoms is described by the Tersoff potential, with the optimized set of parameters obtained by Lindsay and Broido \cite{lindsay_2010} to better describe the thermal properties of graphene. Then, we set two Nosé-Hoover thermostatted regions with a 40~\AA-length near both fixed ends as can be seen in Figure \ref{fig:temperature}. These regions play the role of thermal baths at $T_\text{hot}=500$~K and $T_\text{cold}=300$~K to induce a temperature gradient in the GNRs and, consequently, a heat current that flows from the hot to the cold reservoir. The process of building a temperature gradient is performed by running an NVT MD only for the thermostatted regions and an NVE dynamics for the atoms between the two thermal reservoirs during 1~ns. This ensures that the system reaches a steady state where the energy that enters the cold reservoir is equal in magnitude but opposite in signal to the energy that goes out from the hot one. A similar methodology has been proved to agree well with experimentally-obtained results for pristine graphene~\cite{xu_2014,fan_2017a}.

In Figure \ref{fig:temperature} we present an example of the spatial temperature distribution for a three-knot nanoribbon. Following the equipartition theorem, a 'temperature' per atom is defined as $t = k_\text{at}/k_\text{B}/1.5$ with $k_\text{at}$ the kinetic energy per atom, and $k_\text{B}$ the Boltzmann constant. This is defined in a way that the sum of $t$ over all atoms yields the temperature of the system. The temperature value of each atom in Figure \ref{fig:temperature} is given by the spatial average of all $t$ values of atoms within a cutoff radius of 5~\AA~ as performed by OVITO~\cite{stukowski_2010}. After setting the thermal gradient, we carried out an MD run for another 1~ns to calculate the temperature profile and monitor the cumulative energy that goes in (out) to (from) the cold (hot) reservoir.

The heat current $J$ is defined as the slope of the cumulative energy as a function of time. Using the one-dimensional Fourier's law of heat conduction, the heat flux $q$ can be estimated by
\begin{equation}\label{eq:heatfourier}
    q = \frac{J}{A} = \kappa \frac{dT}{dx},
\end{equation}
where $A$ is the cross-sectional area given by $W*3.35$~\AA, with 3.35~\AA~ being the thickness of a graphene monolayer, $dT/dx$ the temperature gradient, and $\kappa$ the thermal conductivity of the system. 

Discontinuities in the temperature profile of $\Delta T$ characterize a thermal resistance. This concept is usually used when there is an interface between two different materials, which defines the so-called interfacial thermal resistance, or the Kapitza resistance~\cite{pollack_1969} that is given by
\begin{equation}
    R = \frac{\Delta T}{q}.
\end{equation}
Although $R$ is defined for situations when there are two different materials, where this resistance is a result of a heat current flowing through two media with different thermal conductivity, the Kapitza resistance was already reported to exist in structures made out of a single material but with the presence of defects~\cite{wang_2014,medranosandonas_2015,fan_2017,kang_2018} and lattice distortions~\cite{volkov_2012,huang_2011,nishimura_2012,lee_2013,chang_2007}.

\section{Results}

\begin{figure}[!htb]
\centering
\includegraphics[width=\linewidth]{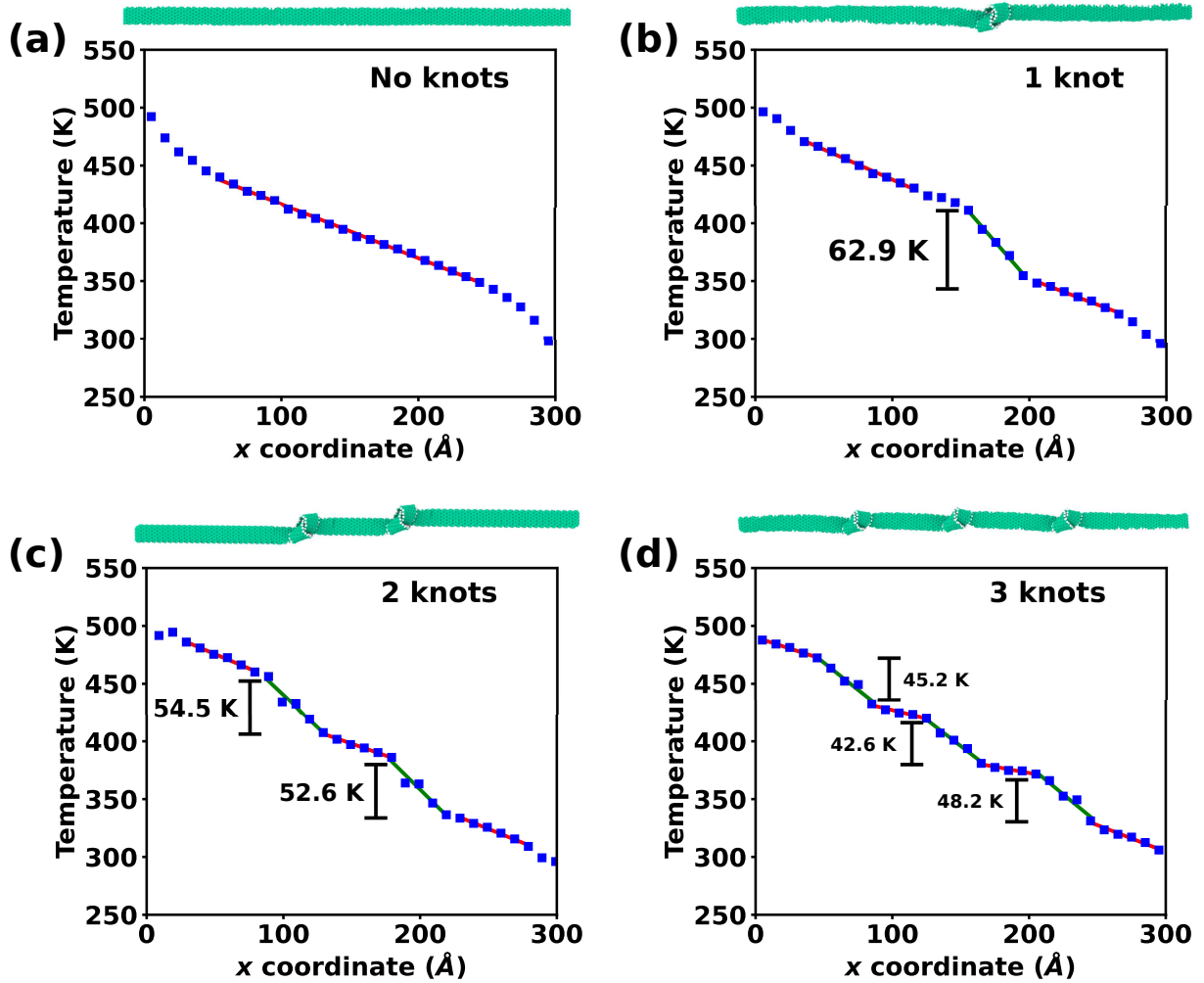}
\caption{Temperature profiles of the cases with (a) zero, (b) one, (c) two, and (d) three knots. Red lines correspond to linear regression lines in regions where there are no knots and a linear behavior between temperature and position (where the Fourier's law still holds). Green lines are the corresponding fittings inside regions where the knots are located. Above each profile, a schematic of the system is displayed to make clearer the effect of each knot. Also, the temperature jumps $\Delta T$ caused by the knots are also highlighted.}
\label{fig:temp_profiles}
\end{figure}

The temperature profile of GNRs without, with one, two, and three knots are presented in Figure \ref{fig:temp_profiles}. The red lines highlight the flat regions in the nanoribbon, where the temperature gradient is approximately linear and have a constant $dT/dx$. The green lines correspond to $dT/dx$ where the knots are located. For the case without knots, a well defined linear behavior can be observed in the Figure \ref{fig:temp_profiles}~(a), where the profile changes exponentially near the thermal baths due to phonon scattering at the interfaces between the regions that are thermostatted and those that are not~\cite{alghalith_2018,allen_2014,cepellotti_2017}. From one to three-knot cases (Figures \ref{fig:temp_profiles}~(b)-(d)), a temperature jump is observed at each knot, where the the slope of the thermal gradient is different from that of the flat regions without knots. Thus, the knots behave as boundary resistances in which the effective thermal conductivity is reduced due to additional phonon scattering at the knot regions.

\begin{figure}[!htb]
\centering
\includegraphics[width=\linewidth]{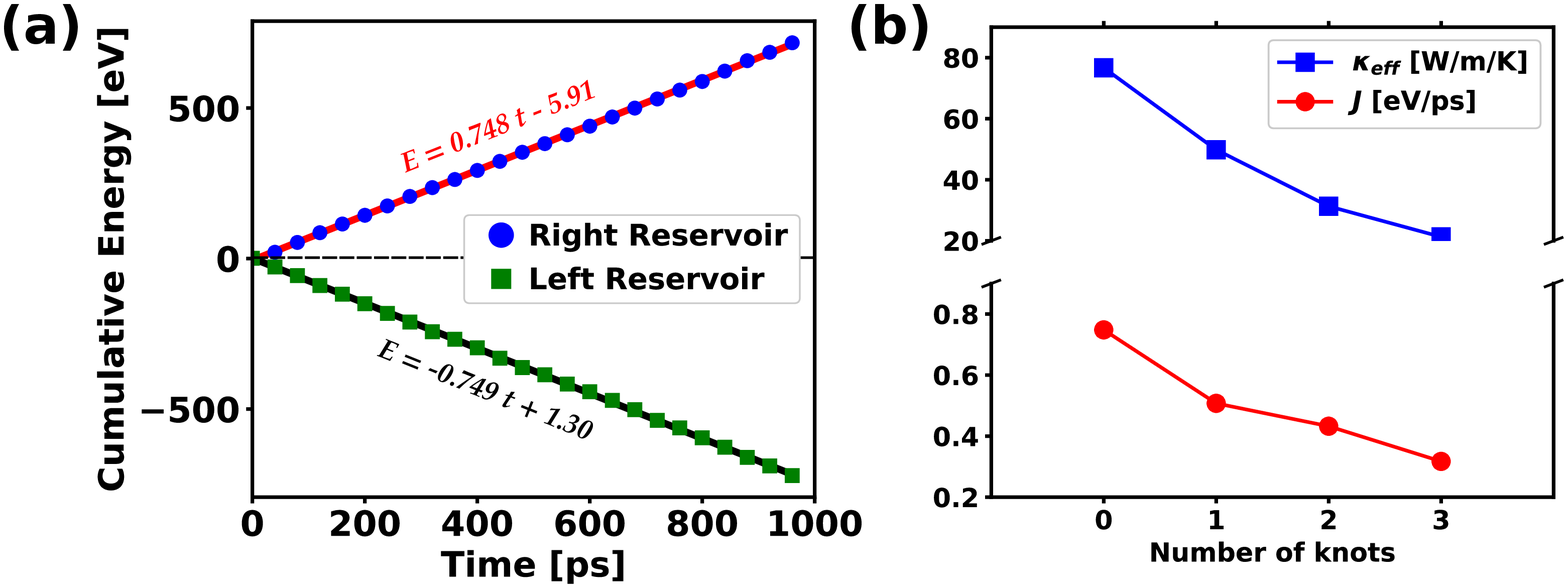}
\caption{Calculation of the heat current. (a) Time evolution of the cumulative energy added/subtracted in the heat baths and their corresponding linear fittings showing the coefficients where the slope allows to estimate the heat current $J$ in each reservoir. The total $J$ is the arithmetic average of the values for each individual reservoir. (b) The behavior of the obtained heat current effective thermal conductivity $\kappa_\text{eff}$ calculated from Equation \ref{eq:effkappa}.}
\label{fig:energy-heat-kappa}
\end{figure}

In order to calculate $\kappa$ (the thermal conductivity of the system), we need to determine the heat flux $q$ produced by NEMD simulations. This can be done by the time evolution of the cumulative energy in both hot and cold reservoirs, where the heat current $J$ is obtained by the average slope from their linear fitting as shown in Figure \ref{fig:energy-heat-kappa}~(a) for the one-knot example. In this case, $J=(0.748 + |-0.749|)/2$. Fourier's law should only be applied to regions where there is a linear relationship between temperature and position. The thermal conductivity of a region is calculated by just dividing the heat flux by the slope of fitted regions. As previously shown in Figure \ref{fig:temp_profiles}, the knots behave as materials with different conductivities $\kappa^k$. The effective thermal conductivity of the whole quasi-1D structures is calculated following the theory of heat transfer in layered composites~\cite{alghalith_2018,badry_2020}, where
\begin{equation}\label{eq:effkappa}
    \frac{1}{\kappa_\text{eff}} = \sum_{i=1}^{N_k+1}\frac{\phi_i}{\kappa_i} + \sum_{i=1}^{N_k}\frac{\phi^k_i}{\kappa^k_i},
\end{equation}
where $N_k$ is the number of knots, $\phi_i=l_i/L_x$ and $\phi^k_i=l^k_i/L_x$ are the length fractions of flat and knotted regions, respectively. $\kappa_i$ and $\kappa^k_i$ are the corresponding thermal conductivities. Each $\kappa_i$ is obtained by
\begin{equation}
    \kappa_i = \frac{q}{\big|\frac{dT}{dx}\big|_i},
\end{equation}
with $q$ being the heat flux in the whole nanoribbon and $|dT/dx|_i$ the slope of the thermal gradient of the region $i$. A similar expression holds with knotted regions. The Equation \ref{eq:effkappa} has already been applied to nanomaterials with geometrical distortions with similar temperature profiles, as those shown in Figure \ref{fig:temp_profiles}. The dependence with the number of knots of both thermal current $J$ and effective thermal conductivity $\kappa_\text{eff}$ are displayed in Figure \ref{fig:energy-heat-kappa}~(b). As each knot introduces an additional thermal resistance in the structure, we observe the expected decay of both $J$ and $\kappa_{eff}$ as $N_k$ increases.

With the heat flux obtained for all cases, we are able to calculate the Kapitza resistance that are induced by the presence of knots in the GNRs. In Table \ref{tab:thermalresistance} we show the temperature jumps $\Delta T$ that appear in each knot for nanoribbons with one, two and three knots. As the number of knots increases, $\Delta T$ decreases but these discontinuities in the temperature profiles have similar values within a given structure. The corresponding values of the resistance $R$ introduced by each knot are almost the same. As the number of knots increases, the heat decreases in the same way to keep the ration $\Delta T / q$ close to a constant value. Although $R$ values for each individual knot are approximately the same, the overall thermal resistance of nanoribbons with two and three knots are, respectively, two and three times higher than that of GNRs with a single knot. This occurs because the resistance of each knot sums up as they are associated in series. 

\begin{table}
\caption{Temperature jumps $\Delta T$ and their corresponding interfacial (Kapitza) thermal resistances $R$ induced by each knot for GNRs with one, two, and three knots.}
\label{tab:thermalresistance}
\begin{center}
\resizebox{\columnwidth}{!}{%
\begin{tabular}{|c|c|c|}\hline
\textbf{Number of knots} & $\boldsymbol{\Delta T}$ \textbf{[K]} & $\boldsymbol{R\times 10^{-10}}$ \textbf{[m$\boldsymbol{^2}$/W$\cdot$K]}  \\ \hline 
\textbf{1} & 62.9 & 2.59  \\ \hline
\textbf{2} & \begin{tabular}{@{}c@{}c@{}} 54.5\\ 52.6\end{tabular} & \begin{tabular}{@{}c@{}c@{}} 2.63\\ 2.54\end{tabular} \\ \hline
\textbf{3} & \begin{tabular}{@{}c@{}c@{}} 45.2 \\ 42.5 \\ 48.2 \end{tabular} & \begin{tabular}{@{}c@{}c@{}}2.98 \\ 2.81 \\ 3.18\end{tabular} \\ \hline
\end{tabular}
}
\end{center}
\end{table}

\begin{figure}[!htb]
\centering
\includegraphics[width=\linewidth]{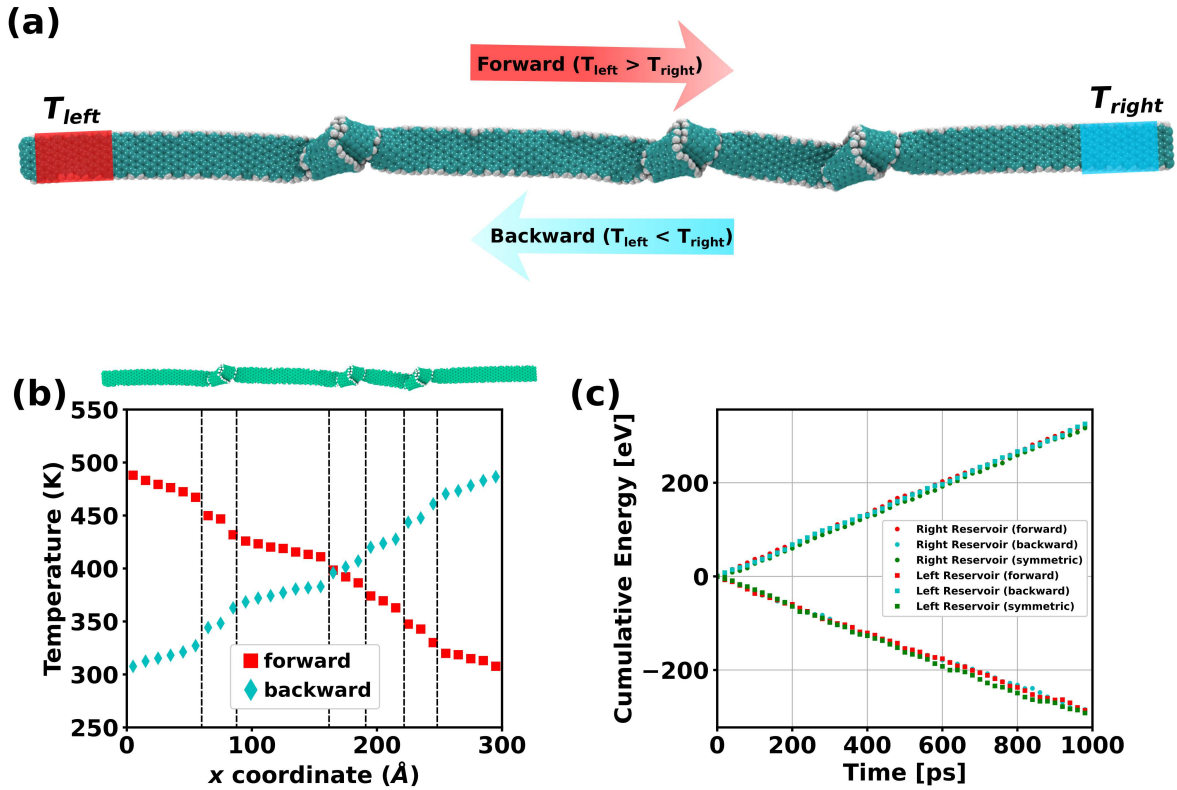}
\caption{Thermal rectification in asymmetrically-placed three-knot GNRs. (a) Simulation setup, where the distance between each successive knot is different. (b) Temperature profile corresponding to the forward (red squares) and backward (cyan diamonds). (c) Time evolution of the cumulative energy subtracted (added) to the hot (cold) reservoirs.}
\label{fig:rectification}
\end{figure}

Lastly, we considered a knotted GNR where one of the knots is displaced relative to the others in order to investigate the thermal transport in asymmetrically placed knots in nanoribbons. This was motivated by previous works~\cite{wang_2014,medranosandonas_2015,cartoix_2015,wang_2017,yang_2020,liu_2021} that thermal rectification appears in asymmetrical graphene and MoS$_2$ nanoribbons. In other words, the presence of asymmetry indicates that when the hot and cold reservoirs are reversed, the induced heat flux should be different. Figure \ref{fig:rectification}~(a) shows the asymmetrical knotted GNR studied here, where the first two knots are 100~\AA~ apart and the last knot is 50~\AA~ distant from the second one. Two regions near the fixed ends of the structure are thermostatted at temperatures $T_\text{left}$ and $T_\text{right}$. The heat flux vector points in the forward (backward) direction when $T_\text{left}>T_\text{right}$ ($T_\text{left}<T_\text{right}$). Figure \ref{fig:rectification}~(b) shows the temperature profiles of both forward and backward cases. The profile of the backward case is essentially a mirror inversion of the forward case, showing that they are likely to present the same conductivity. The accumulated energy for GNRs with three knots is shown in Figure \ref{fig:rectification}~(c). We compared cases with a symmetrical arrangement of the three knots and the asymmetrical cases (forward and backward), and we observed that the cumulative energy at both heat reservoirs for all cases possess almost the same value after running for 1~ns, in which this difference ($\sim 0.7\%$) is close to the error to reach the steady state ($\sim 0.5\%$). To quantify the thermal rectification, we computed the rectification factor
\begin{equation}
    \eta = \frac{q_f}{q_b} - 1,
\end{equation}
where $q_f$ ($q_b$) is the heat flux in the forward (backward) direction. For our case of asymmetrically arranged knots, we obtain $\eta \sim 0.3 \%$. For comparison, rectification factors computed with NEMD for asymmetrical GNRs and MoS$_2$ range from $1-25\%$~\cite{wang_2014,medranosandonas_2015,wang_2017}, showing that for the cases considered here, the rectification in knotted GNRs does not reach values large enough to consider them as thermal diodes.

\section{Conclusions}\label{sec:conclusion}

We studied the effect of structural knots in the thermal transport of graphene nanoribbons using non-equilibrium molecular dynamics simulations. The influence of the number of knots on the heat current was investigated by considering cases where one, two, and three knots. We observed that there is a temperature jump at the position of each knot, which is an effect similar to interfacial Kapitza resistance present between two different materials and/or single materials with defects and/or lattice distortions. Our study also demonstrates that in the three-knot nanoribbon case, the heat current is insensitive to the relative position of the middle knot and the direction of the energy flux through the system. Thus, for the cases considered here, the effect of thermal rectification is not significant.

\section*{Acknowledgments}
This work was financed by Sao Paulo State Research Support Foundation (FAPESP). We also thank the Center for Computing in Engineering and Sciences at Unicamp for financial support through the FAPESP/CEPID Grants \#2013/08293-7 and \#2018/11352-7. 







\end{document}